\documentclass[a4paper,11pt]{article}%
\usepackage{geometry}
\usepackage{amsmath}
\usepackage{amsfonts}
\usepackage{amssymb}
\usepackage{graphicx}
\usepackage{indentfirst}
\usepackage[small,bf]{caption}
\usepackage{slashed}
\providecommand{\U}[1]{\protect\rule{.1in}{.1in}}

\begin{document}

\date{}
\title{\textbf{Study of the all orders multiplicative renormalizability of a local confining quark action in the Landau gauge}}
\author{\textbf{M.~A.~L.~Capri}\thanks{caprimarcio@gmail.com}\,\,,
\textbf{D.~Fiorentini }\thanks{diegofiorentinia@gmail.com}\,\,,
\textbf{S.~P.~Sorella}\thanks{silvio.sorella@gmail.com}\,\,\,,\\[2mm]
{\small \textnormal{  \it Departamento de F\'{\i }sica Te\'{o}rica, Instituto de F\'{\i }sica, UERJ - Universidade do Estado do Rio de Janeiro,}}
 \\ \small \textnormal{ \it Rua S\~{a}o Francisco Xavier 524, 20550-013 Maracan\~{a}, Rio de Janeiro, Brasil}\normalsize}
\maketitle

\begin{abstract}
The inverse of the Faddeev-Popov operator plays a pivotal role within the Gribov-Zwanziger approach to the quantization of Euclidean Yang-Mills theories in Landau gauge. Following a recent proposal \cite{Capri:2014bsa}, we show  that the inverse of the  Faddeev-Popov operator can be consistently coupled to quark fields. Such a coupling gives rise to a local action while reproducing the behaviour of the quark propagator observed in lattice numerical simulations in the non-perturbative infrared region. By using the algebraic renormalization framework, we prove that the aforementioned local action is multiplicatively renormalizable to all orders.

\end{abstract}

\section{Introduction}

Nowadays, the so called Gribov-Zwanziger framework \cite{Gribov:1977wm,Zwanziger:1988jt,Zwanziger:1989mf,Zwanziger:1992qr}   is a powerful tool in order to study non-perturbative aspects  of  gluon and quark confinement. Going a step beyond of the perturbative quantization method of Faddeev-Popov, Gribov called attention to the fact that the presence of zero modes of the Faddeev-Popov operator, given by
\begin{equation}
\mathcal{M}^{ab}\, =-\partial_{\mu}(\delta^{ab}\partial_{\mu}\,-gf^{abc}A^{c}_{\mu} )\,\,,
\label{FP_op}
\end{equation}
results in the existence of  Gribov copies \cite{Gribov:1977wm}, {\it i.e.}  equivalent gauge field configurations fulfilling  the same gauge-fixing condition\footnote{For a pedagogical introduction to the Gribov problem, see \cite{Sobreiro:2005ec,Vandersickel:2012tz}}. As a consequence, the Faddeev-Popov quantization procedure becomes ill-defined, meaning that it is not possible to pick up a unique gauge field configuration for each gauge orbit through a local and covariant gauge-fixing condition \cite{Singer:1978dk}. The existence of the Gribov copies is a non-perturbative phenomenon which has deep consequences on the infrared behaviour of confining Yang-Mills theories. \\\\To deal with the existence of equivalent gauge field configurations, Gribov was able to show that a large number of copies could be eliminated by restricting the domain of integration in the functional integral to a certain region $\Omega$ in field space \cite{Gribov:1977wm}. This region is known as the Gribov region and is given by all field configurations which fulfil the Landau gauge condition, $\partial_\mu A^{a}_\mu = 0$, and for which the Faddeev-Popov operator $\mathcal{M}^{ab}$ is strictly positive, namely
\begin{equation}
\Omega \;= \; \{ A^a_{\mu}\;; \;\; \partial_\mu A^a_{\mu}=0\;; \;\; {\cal M}^{ab}=-(\partial^2 \delta^{ab} -g f^{abc}A^{c}_{\mu}\partial_{\mu})\; >0 \; \} \;. \label{gr}
\end{equation}
The region $\Omega$ enjoys important properties \cite{Dell'Antonio:1989jn,Dell'Antonio:1991xt,vanBaal:1991zw}:
\begin{itemize}
\item  {\it i)} it is convex and bounded in all directions in field space. Its boundary $\partial \Omega$ is the first Gribov horizon, where the first non vanishing eigenvalue of  $\mathcal{M}^{ab}$ shows up.
\item  {\it  ii)}  every gauge orbit intersects at least once the region $\Omega$. This last property gives a well defined support to Gribov's original proposal of cutting off the functional integral at the Gribov horizon.
\end{itemize}
Later on, Zwanziger \cite{Zwanziger:1988jt,Zwanziger:1989mf,Zwanziger:1992qr}  showed that the restriction of the domain of integration to the region $\Omega$  is equivalent to adding to the original Faddeev-Popov action a nonlocal term $H(A)$, called horizon function, given by
\begin{equation}
H(A)=g^{2}\int d^{4}xd^{4}y\,f^{abc}A^{b}_{\mu}\; (\mathcal{M}^{-1})^{ad}(x,y)f^{dec}\; A^{e}_{\mu}(y)\,,
\label{horizon_function}
\end{equation}
with $(\mathcal{M}^{-1})^{ab}$ being the inverse of the Faddeev-Popov operator \eqref{FP_op}. \\\\Thus, for the partition function one writes \cite{Gribov:1977wm,Zwanziger:1988jt,Zwanziger:1989mf,Zwanziger:1992qr}
\begin{eqnarray}
 Z  &= &  \;
\int_\Omega {\cal D}A\; \delta(\partial A)\; det( {\cal M}) \;e^{-S_{\mathrm{YM}}}  =   \int {\cal D}A\;  \delta(\partial A)\; det( {\cal M}) \; e^{-(S_{\mathrm{YM} }+\gamma^4 H(A) -V\gamma^4 4(N^2-1))}  \nonumber  \\
& = & \int {\cal D}A\;  {\cal D}c\;{\cal D}\bar{c}\; {\cal D} b \; e^{-S_{\mathrm{GZ} }}
\;, \label{zww1}
\end{eqnarray}
where  $S_{\mathrm{GZ} }$ is the  Gribov-Zwanziger action
\begin{equation}
S_{\mathrm{GZ}}=S_{\mathrm{FP}}+\gamma^{4}H(A)  -V\gamma^4 4(N^2-1)   \,,
\label{GZ_action}
\end{equation}
with $S_{\mathrm{FP}}$ denoting the Faddeev-Popov action in Landau gauge
\begin{equation}
S_{\mathrm{FP}} = \int d^{4}x \left(\frac{1}{4}F^{a}_{\mu\nu}F^{a}_{\mu\nu}
+b^{a}\,\partial_{\mu}A^{a}_{\mu}-\bar{c}^{a}\mathcal{M}^{ab}c^{b}\right)   \;, \label{sfp}
\end{equation}
The field $b^{a}$ in expression \eqref{sfp} is the Lagrange multiplier enforcing the Landau gauge condition, $\partial_\mu A^a_\mu=0$, while  $(c^{a},\bar{c}^{a})$ are the Faddeev-Popov ghosts. Also, the massive parameter  $\gamma^{2}$ in eq.\eqref{GZ_action} is the so called Gribov parameter. It is not a free parameter, being determined in a self-consistent way through a gap equation, called the horizon condition, which reads
\begin{equation}
 \frac{\partial \mathcal{E}_v}{\partial\gamma^2}=0\;,   \label{ggap}
\end{equation}
where $\mathcal{E}_{v}(\gamma)$ is the vacuum energy defined by
\begin{equation}
 e^{-V\mathcal{E}_{v}}=\;Z\;  \label{vce} \;.
\end{equation}
It is worth to point out that, even if the horizon term  $H(A)$, eq.\eqref{horizon_function}, is non-local, the action \eqref{GZ_action}  can be cast in local form \cite{Zwanziger:1988jt,Zwanziger:1989mf,Zwanziger:1992qr} by means of the introduction of a set of auxiliary fields  $(\bar{\omega}_\mu^{ab}, \omega_\mu^{ab}, \bar{\varphi}_\mu^{ab},\varphi_\mu^{ab})$, where $(\bar{\varphi}_\mu^{ab},\varphi_\mu^{ab})$ are a pair of bosonic fields, while $(\bar{\omega}_\mu^{ab}, \omega_\mu^{ab})$ are anti-commuting. For the local formulation of the theory, we have
\begin{equation}
 Z = \;
\int {\cal D}A\;{\cal D}c\;{\cal D}\bar{c}\; {\cal D} b \; {\cal D}{\bar \omega}\; {\cal D} \omega\; {\cal D} {\bar \varphi} \;{\cal D} \varphi \; e^{-S_{\mathrm{GZ}}} \;, \label{lzww1}
\end{equation}
where $S_{\mathrm{GZ}}$ is now given by the local expression
\begin{equation}
S_{\mathrm{GZ}} =  S_{\mathrm{FP}} + S_0 + S_\gamma  \;, \label{sgz}
\end{equation}
where  $S_0$ and $S_\gamma$ read
\begin{equation}
S_0 =\int d^{4}x \left( {\bar \varphi}^{ac}_{\mu} (\partial_\nu D^{ab}_{\nu} ) \varphi^{bc}_{\mu} - {\bar \omega}^{ac}_{\mu}  (\partial_\nu D^{ab}_{\nu} ) \omega^{bc}_{\mu}  - gf^{amb} (\partial_\nu  {\bar \omega}^{ac}_{\mu} ) (D^{mp}_{\nu}c^p) \varphi^{bc}_{\mu}  \right) \;, \label{s0}
\end{equation}
and
\begin{equation}
S_\gamma =\; \gamma^{2} \int d^{4}x \left( gf^{abc}A^{a}_{\mu}(\varphi^{bc}_{\mu} + {\bar \varphi}^{bc}_{\mu})\right)-4 \gamma^4V (N^2-1)\;. \label{hfl}
\end{equation}
The local action \eqref{sgz} exhibits the important property of being multiplicative renormalizable to all orders \cite{Zwanziger:1988jt,Zwanziger:1989mf,Zwanziger:1992qr}. \\\\Recently, a refinement of the Gribov-Zwanziger action has been worked out by the authors  \cite{Dudal:2007cw,Dudal:2008sp,Dudal:2011gd}, by taking into account the existence of  dimension two condensates.   The Refined Gribov-Zwanziger (RGZ) action reads \cite{Dudal:2007cw,Dudal:2008sp,Dudal:2011gd}
\begin{equation}
S_{\mathrm{RGZ}} = S_{\mathrm{GZ}} + \int d^4x \left(  \frac{m^2}{2} A^a_\mu A^a_\mu  - \mu^2 \left( {\bar \varphi}^{ab}_{\mu}  { \varphi}^{ab}_{\mu} -  {\bar \omega}^{ab}_{\mu}  { \omega}^{ab}_{\mu} \right)   \right)  \;.  \label{rgz}
\end{equation}
As the Gribov parameter $\gamma^2$, the massive parameters $(m^2, \mu^2)$ have a dynamical origin, being related to the existence of the dimension two condensates $\langle A^a_\mu A^a_\mu \rangle$ and  $\langle {\bar \varphi}^{ab}_{\mu}  { \varphi}^{ab}_{\mu} -  {\bar \omega}^{ab}_{\mu}  { \omega}^{ab}_{\mu}  \rangle$, \cite{Dudal:2007cw,Dudal:2008sp,Dudal:2011gd}. The gluon propagator obtained from the RGZ action turns out to be suppressed in the infrared region, attaining a non-vanishing value at zero momentum, $k^2=0$, {\it i.e.}
\begin{eqnarray}
\langle  A^a_\mu(k)  A^b_\nu(-k) \rangle  & = &  \delta^{ab}  \left(\delta_{\mu\nu} - \frac{k_\mu k_\nu}{k^2}     \right)   {\cal D}(k^2) \;, \label{glrgz} \\
{\cal D}(k^2) & = & \frac{k^2 +\mu^2}{k^4 + (\mu^2+m^2)k^2 + 2Ng^2\gamma^4 + \mu^2 m^2}  \;. \label{Dg}
\end{eqnarray}
Also, the ghost propagator stemming from the Refined theory is not enhanced in the deep infrared
\begin{equation}
{\cal G}^{ab}(k^2) = \langle  {\bar c}^{a} (k)  c^b(-k) \rangle \Big|_{k\sim 0} \; \sim \frac{\delta^{ab}}{k^2}   \;.\label{ghrgz}
\end{equation}
The infrared behaviour of the  gluon and ghost propagators obtained from the RGZ  action turns out to be in very good agreement with the most recent  numerical lattice simulations on large lattices \cite{Cucchieri:2007rg,Cucchieri:2008fc,Cucchieri:2011ig,Oliveira:2012eh,Oliveira:2008uf}. Furthermore, from  the numerical estimates  \cite{Cucchieri:2011ig}  of the parameters $(m^2,\mu^2,\gamma^2)$ it turns out that  the RGZ gluon propagator \eqref{glrgz} displays complex poles and violates  reflection positivity. This kind of propagator lacks the  K{\"a}ll{\'e}n-Lehmann spectral representation and cannot be associated with the propagation of physical particles. Rather, it indicates that, in the non-perturbative infrared region, gluons are not physical excitations of the spectrum of the theory, {\it i.e.} they are confined. Let us mentioning here that the RGZ gluon propagator has been employed in analytic calculation of glueball states  \cite{Dudal:2010cd,Dudal:2013wja}, yielding results which compare well with the available numerical simulations as well as with other approaches, see \cite{Mathieu:2008me} for an  account on this topic. \\\\As illustrated above, the RGZ framework, turns out to capture important aspects of the  gluon confinement. Nevertheless, what can be said about the matter sector? A first answer to this question was proposed  recently in \cite{Capri:2014bsa} where a new term, very similar to  Zwanziger's horizon function, was investigated. Such new term was inspired by  recent lattice numerical simulations  \cite{Cucchieri:2014via} of the correlation function
\begin{equation}
Q^{abcd}_{\mu\nu}(x-y)=\left\langle \mathcal{R}^{ab}_{\mu}(x)\mathcal{R}^{cd}_{\nu}(y)\right\rangle\,,   \label{qabcd}
\end{equation}
with
\begin{equation}
\mathcal{R}^{ac}_{\mu}(x)=\int d^4z(\mathcal{M}^{-1})^{ad}(x,z)gf^{dec}A^{e}_{\mu}(z)\,,
\label{R}
\end{equation}
More precisely, in  \cite{Cucchieri:2014via}, it has been shown that the Fourier transform of expression \eqref{qabcd}  is non-vanishing and behaves as $\frac{1}{k^4}$ in the deep infrared, a result which is again in agreement with the RGZ framework, {\it i.e.}
\begin{equation}
 \langle {\tilde {\cal R}} ^{ab}_\mu(k)  {\tilde {\cal R}}^{cd}_\nu(-k)  \rangle\Big|_{k\sim 0}   \sim \frac{1}{k^4} \;.  \label{enhanc}
\end{equation}
As observed in \cite{Cucchieri:2014via}, this behaviour can be understood by making use of the analysis of \cite{Zwanziger:2010iz}, {\it i.e.} of the cluster decomposition
\begin{equation}
 \langle {\tilde {\cal R}} ^{ab}_\mu(k)  {\tilde {\cal R}}^{cd}_\nu(-k)  \rangle    \sim  g^2 {\cal G}^2(k^2) {\cal D}(k^2) \;, \label{clust}
\end{equation}
where ${\cal D}(k^2)$ and ${\cal G}(k^2)$ correspond to the   gluon and ghost propagators, eqs.\eqref{Dg},\eqref{ghrgz}. A non-enhanced ghost propagator, {\it i.e.}  ${\cal G}(k^2) \Big|_{k\sim 0} \sim \frac{1}{k^2}$, and an infrared finite gluon propagator, {\it i.e.} ${\cal D}(0) \neq 0$, nicely yield the behaviour of eq.\eqref{enhanc}. \\\\In \cite{Capri:2014bsa}, the idea that the quantity $\mathcal{R}^{ab}_{\mu}(x)$ and the correlation function $Q^{abcd}_{\mu\nu}(x-y)$, eqs.\eqref{qabcd}, \eqref{R},  could be generalized to the case of matter fields was exploited in details. The main argument developed in \cite{Capri:2014bsa} can be summarized as follows.  Let $\mathfrak{F}^{\,i}(x)$ be a generic matter field,  {\it i.e.} a scalar or a spinor field, in a given representation of the gauge group $SU(N)$, and let $\mathcal{R}_{\mathfrak{F}}^{ai}(x)$ and its complex conjugate  $\bar{\mathcal{R}}_{\mathfrak{F}}^{ai}(x)$ be defined by
\begin{eqnarray}
\mathcal{R}_{\mathfrak{F}}^{ai}(x)&:=&g\int d^{4}z (\mathcal{M}^{-1})^{ab}(x,z)(T^{b})^{ij}\mathfrak{F}^{j}(z)\,,\nonumber\\
\bar{\mathcal{R}}_{\mathfrak{F}}^{ai}(x)&:=&g\int d^{4}z (\mathcal{M}^{-1})^{ab}(x,z)\bar{\mathfrak{F}}^{j}(z)(T^{b})^{ji}\,,
\label{R_F}
\end{eqnarray}
where $(T^{a})^{ij}$ are the generators of the representation and $a=1,\dots,(N^{2}-1)$. The quantity $\bar{\mathcal{R}}_{\mathfrak{F}}^{ai}(x)$ is needed if the generic field $\mathfrak{F}^{i}$ is complex. Notice how similar  eq's \eqref{R} and \eqref{R_F} are. Then, a non-trivial correlation function
\begin{equation}
Q^{abij}_{\mathfrak{F}}(x-y)=\left\langle {\mathcal{R}}^{ai}_{\mathfrak{F}}(x)\bar{\mathcal{R}}^{bj}_{\mathfrak{F}}(y)\right\rangle\,,
\end{equation}
can be obtained from a theory constructed by adding to the usual matter action a term similar to the horizon function, namely
\begin{equation}
S_{\mathrm{matter}}\to S_{\mathrm{matter}}+M^{3}H_{\mathrm{matter}}\,,
\label{matter_action_F}
\end{equation}
where, in complete analogy with the horizon function \eqref{horizon_function}, $H_{\mathrm{matter}}$ is given by
\begin{equation}
H_{\mathrm{matter}}=-g^2\int d^4xd^4y\,\bar{\mathfrak{F}}^{\,i}(x)(T^a)^{ij}(\mathcal{M}^{-1})^{ab}(x,y)(T^b)^{jk}\mathfrak{F}^{\,k}(y)  \;.
\label{horizon_function_matter}
\end{equation}
Also, the mass parameter\footnote{In the case in which the field $\mathfrak{F}^{\,i}$ is a scalar field, the power of the mass parameter $M$ appearing in expression \eqref{matter_action_F} is four, due to the fact that a scalar field has dimension one \cite{Capri:2014bsa}. }  $M$ plays a role akin to that of the Gribov parameter $\gamma$. \\\\The present work aims at pursuing the analysis started in \cite{Capri:2014bsa}. In particular, we shall prove the multiplicative renormalizability of the model constructed according to \eqref{matter_action_F} when the matter field under consideration is the quark field, {\it i.e.}, when $\mathfrak{F}^{\,i}\equiv\psi^{i}$.  It is worth underlining that this approach provides a tree level quark propagator that can be written as
\begin{equation}
\left\langle{\psi}^{i}(p)\bar{\psi}^{j}(-p)\right\rangle=\frac{-ip_{\mu}\gamma_{\mu}+\mathcal{A}(p^{2})}{p^{2}+\mathcal{A}^2(p^{2})}\delta^{ij}\,,
\label{fermionic_propagator}
\end{equation}
where the quark mass function $\mathcal{A}(p^{2})$ is given by
\begin{equation}
\mathcal{A}(p^{2})=m_{\psi}+g^{2}\left(\frac{N^{2}-1}{2N}\right)\,\frac{M^{3}}{p^{2}+\mu^{2}_{\psi}}\,.   \label{aq}
\end{equation}
Here, the mass $m_{\psi}$ stands for the quark mass, while $\mu^{2}_{\psi}$ is related  to the condensation of a lower  dimensional operator in the matter
sector \cite{Capri:2014bsa}, a feature which shares great similarity with the lower dimensional condensates of the RGZ action, eq.\eqref{rgz}.  \\\\It is worth underlining here that a quark propagator of the kind of eqs.\eqref{fermionic_propagator}, \eqref{aq} fits very well the lattice numerical  data, see  \cite{Parappilly:2005ei} and the discussion reported in \cite{Capri:2014bsa}. In particular, one notices that the quark mass function does not vanish at zero momentum for vanishing  quark mass $m_{\psi}=0$, signalling a dynamical breakdown of the chiral symmetry.  As such, expressions \eqref{fermionic_propagator},\eqref{aq} capture nontrivial non-perturbative aspects of quark confinement. \\\\The present work  is organized as follows. In Section 2 we construct the complete local action in the matter sector which takes into account the term \eqref{horizon_function_matter}. In Section 3  we establish the large set of Ward identities fulfilled by the local action. In Section 4 we characterize  the most general local invariant counterterm by means of the algebraic renornalization procedure \cite{Piguet:1995er} and we establish the all order multiplicative renormalizability of the model. In Section 5 we collect our conclusions.

\section{Identifying the local classical starting action}
In order to identify the local classical action implementing the framework discussed above, we start with the Gribov-Zwanziger action supplemented with a non-local matter term, as described in eq.\eqref{horizon_function_matter}, {\it i.e.} 
\begin{equation}
S=S_{\mathrm{GZ}}+S_{\mathrm{matter}}+ M^3 H_{\mathrm{matter}} \,,
\end{equation}
where $S_{\mathrm{GZ}}$ is the Gribov-Zwanziger action, eq. \eqref{sgz}, and $S_{\mathrm{matter}}, H_{\mathrm{matter}}$  given, respectively, by  
\begin{equation}
S_{\mathrm{matter}}=\int d^{4}x\,\Bigl[\bar{\psi}^{i\alpha}(\gamma_{\mu})_{\alpha\beta}D^{ij}_{\mu}\psi^{j\beta}
-m_{\psi}\bar{\psi}^{i}_{\alpha}\psi^{i\alpha}\Bigr]   \;, 
\label{usualmatter}
\end{equation}
and
\begin{equation}
H_{\mathrm{matter}}=-g^{2}\int d^{4}x\,d^{4}y\,\bar\psi^{i}_{\alpha}(x) (T^{a})^{ij}(\mathcal{M}^{-1})^{ab}(x,y)(T^{b})^{jk}\psi^{k\alpha}(y)\,.
\label{nonlocal}
\end{equation}
Despite its non-locality, the term $H_{\mathrm{matter}}$  can be cast in local by means of the introduction of a  set of localizing field, in a way similar to the localization of the horizon function \eqref{horizon_function} \cite{Zwanziger:1988jt,Zwanziger:1989mf,Zwanziger:1992qr}. Therefore, for the local version of the starting action, we write 
\begin{equation}
S=S_{\mathrm{GZ}}+S_{\mathrm{matter}}^{\mathrm{local}}   \label{starting} 
\end{equation}
where 
\begin{eqnarray}
S_{\mathrm{matter}}^{\mathrm{local}}&=&\int d^{4}x\,\Bigl[\bar{\psi}^{i\alpha}(\gamma_{\mu})_{\alpha\beta}D^{ij}_{\mu}\psi^{j\beta}
-m_{\psi}\bar{\psi}^{i}_{\alpha}\psi^{i\alpha}\Bigr]\nonumber\\
&&+\int d^{4}x\,
\left[+\bar{\lambda}^{ai}_{\alpha}(-\partial_{\mu}D^{ab}_{\mu})\lambda^{bi\alpha}+\bar\eta^{ai}_{\alpha}(-\partial_{\mu}D^{ab}_{\mu})\eta^{bi\alpha}
-gf^{abc}(\partial_{\mu}\bar\eta^{ai}_{\alpha})(D^{bd}_{\mu}c^{d})\lambda^{ci\alpha}\right]\nonumber\\
&&+gM^{3/2}\int d^{4}x\,\Bigl[\bar\lambda^{ai}_{\alpha}(T^{a})^{ij}\psi^{j\alpha}
+\bar\psi^{i}_{\alpha}(T^{a})^{ij}\lambda^{aj\alpha}\Bigr]\,,
\end{eqnarray}
where $(\bar\lambda^{ai}, \lambda^{ai})$ are anti-commutating spinor fields and $(\bar\eta^{ai},\eta^{ai})$ are commutating ones. It is easily checked that integration over the auxiliary fields  $(\bar\lambda^{ai}, \lambda^{ai}, \bar\eta^{ai},\eta^{ai})$ gives back the non-local expression  \eqref{nonlocal}. \\\\As  in the case of the Gribov-Zwanziger action \cite{Capri:2014bsa,Zwanziger:1988jt,Zwanziger:1989mf,Zwanziger:1992qr,Dudal:2008sp,Baulieu:2008fy,Dudal:2009xh,Sorella:2009vt,Capri:2010hb,Dudal:2012sb,Reshetnyak:2013bga},  the local action $S$ exhibits a soft breaking of the BRST symmetry, namely 
\begin{equation}
sS=\gamma^{2}\Delta_{\gamma}+M^{3/2}\Delta_{M}\,,
\end{equation}
with
\begin{eqnarray}
\Delta_{\gamma}&=&\int d^{4}x\left[-gf^{abc}(D^{ad}_{\mu}c^{d})(\varphi^{bc}_{\mu}+\bar\varphi^{bc}_{\mu})
+gf^{abc}A^{a}_{\mu}\omega^{bc}_{\mu}\right]\,,\\
\Delta_{M}&=&\int d^{4}x\left[ig^{2}(T^{a})^{ij}(T^{b})^{jk}\bar\lambda^{ai}_{\alpha}c^{b}\psi^{k\alpha}
-ig^{2}(T^{b})^{ki}(T^{a})^{ij}\bar\psi^{k}_{\alpha}c^{b}\lambda^{aj\alpha}
-g(T^{a})^{ij}\bar\psi^{i}_{\alpha}\eta^{aj\alpha}\right]\,,
\end{eqnarray}
and $s$ denoting the nilpotent BRST transformations
\begin{eqnarray}
&\displaystyle sA^{a}_{\mu}=-D^{ab}_{\mu}c^{b}\,,\qquad sc^{a}=\frac{g}{2}f^{abc}c^{b}c^{c}\,,\qquad
s\bar{c}^{a}=b^{a}\,,\qquad
sb^{a}=0\,,&\nonumber\\\nonumber\\
&s\bar\omega^{ab}_{\mu}=\bar\varphi^{ab}_{\mu}\,,\qquad s\bar\varphi^{ab}_{\mu}=0\,,\qquad
s\varphi^{ab}_{\mu}=\omega^{ab}_{\mu}\,,\qquad s\omega^{ab}_{\mu}=0\,,&\nonumber\\\nonumber\\
&s\psi^{i}_{\alpha}=-ig(T^{a})^{ij}\,c^{a}\psi^{j}_{\alpha}\,,\qquad
s\bar\psi^{i}_{\alpha}=-ig\,\bar\psi^{j}_{\alpha}(T^{a})^{ji}c^{a}\,,&\nonumber\\\nonumber\\
&s\bar\eta^{ai}_{\alpha}=\bar\lambda^{ai}_{\alpha}\,,\qquad s\bar\lambda^{ai}_{\alpha}=0\,,\qquad
s\lambda^{ai}_{\alpha}=\eta^{ai}_{\alpha}\,,\qquad s\eta^{ai}_{\alpha}=0\,.      \label{brstf} &
\end{eqnarray}
Notice that, being of dimensions less than four in the fields, the breaking terms $\Delta_{\gamma}, \Delta_{M}$ are soft. This kind of breaking can be kept under control through the renormalization process. To that purpose, we follow the strategy already employed in the case of the GZ and RGZ actions \cite{Zwanziger:1988jt,Zwanziger:1989mf,Zwanziger:1992qr,Dudal:2008sp,Baulieu:2008fy,Dudal:2009xh,Sorella:2009vt,Capri:2010hb,Dudal:2012sb} and we embed the action $S$ into a larger action exhibiting exact BRST invariance. \\\\Following \cite{Zwanziger:1988jt,Zwanziger:1989mf,Zwanziger:1992qr,Dudal:2008sp,Baulieu:2008fy,Dudal:2009xh,Sorella:2009vt,Capri:2010hb,Dudal:2012sb},  let us introduce two set of BRST quartet of external sources:
\begin{equation}
s\bar{N}^{ab}_{\mu\nu}=\bar{M}^{ab}_{\mu\nu}\,,\qquad
s\bar{M}^{ab}_{\mu\nu}=0\,,\qquad
sM^{ab}_{\mu\nu}=N^{ab}_{\mu\nu}\,,\qquad
sN^{ab}_{\mu\nu}=0\,;
\end{equation}
\begin{equation}
s\bar{U}^{ij}_{\alpha\beta}=\bar{V}^{ij}_{\alpha\beta}\,,\qquad s\bar{V}^{ij}_{\alpha\beta}=0\,,\qquad
sV^{ij}_{\alpha\beta}=U^{ij}_{\alpha\beta}\,,\qquad sU^{ij}_{\alpha\beta}=0\,.
\end{equation}
Therefore, we  replace the action $S$ by the BRST invariant action 
\begin{eqnarray}
S_{\mathrm{inv}}&=&\int d^{4}x\,\left[\frac{1}{4}F^{a}_{\mu\nu}F^{a}_{\mu\nu}
+\bar{\psi}^{i\alpha}(\gamma_{\mu})_{\alpha\beta}D^{ij}_{\mu}\psi^{j\beta}
-m_{\psi}\bar{\psi}^{i}_{\alpha}\psi^{i\alpha}\right]\nonumber\\
&&+s\int d^{4}x\,\Bigl(\bar{c}^{a}\,\partial_{\mu}A^{a}_{\mu}
+\bar\omega^{ac}_{\nu}\,\partial_{\mu}D^{ab}_{\mu}\varphi^{bc}_{\nu}
-\bar{N}^{ac}_{\mu\nu}\,D^{ab}_{\mu}\varphi^{bc}_{\nu}
-M^{ac}_{\mu\nu}\,D^{ab}_{\mu}\bar\omega^{bc}_{\nu} - \bar{N}^{ab}_{\mu\nu}M^{ab}_{\mu\nu}\Bigr)\nonumber\\
&&+s\int d^{4}x\,\left[\bar\eta^{ai}_{\alpha}(-\partial_{\mu}D^{ab}_{\mu})\lambda^{bi\alpha}+
\bar{U}^{jk}_{\alpha\beta}\,\bar\psi^{i\alpha}g(T^{a})^{ij}\lambda^{ak\beta}
+V^{jk}_{\alpha\beta}\,\bar\eta^{ak\beta}g(T^{a})^{ij}\psi^{j\alpha}
+\zeta m_{\psi}\,\bar{U}^{ij}_{\alpha\beta}V^{ij}_{\alpha\beta}\right]\nonumber\\
&=&\int d^{4}x\,\biggl\{\frac{1}{4}F^{a}_{\mu\nu}F^{a}_{\mu\nu}
+\bar{\psi}^{i\alpha}(\gamma_{\mu})_{\alpha\beta}D^{ij}_{\mu}\psi^{j\beta}
-m_{\psi}\bar{\psi}^{i}_{\alpha}\psi^{i\alpha}+b^{a}\,\partial_{\mu}A^{a}_{\mu}
+\bar{c}^{a}\,\partial_{\mu}D^{ab}_{\mu}c^{b}\nonumber\\
&&+\bar\varphi^{ac}_{\nu}\,\partial_{\mu}D^{ab}_{\mu}\varphi^{bc}_{\nu}
-\bar\omega^{ac}_{\nu}\,\partial_{\mu}D^{ab}_{\mu}\omega^{bc}_{\nu}
-gf^{abc}(\partial_{\mu}\bar\omega^{ae}_{\nu})(D^{bd}_{\mu}c^{d})\varphi^{ce}_{\nu}
-\bar{M}^{ac}_{\mu\nu}\,D^{ab}_{\mu}\varphi^{bc}_{\nu}\nonumber\\
&&
+\bar{N}^{ac}_{\mu\nu}\left[D^{ab}_{\mu}\omega^{bc}_{\nu}+gf^{abd}(D^{de}_{\mu}c^{e})\varphi^{bc}_{\nu}\right]
-M^{ac}_{\mu\nu}\left[D^{ab}_{\mu}\bar\varphi^{bc}_{\nu}+gf^{abd}(D^{de}_{\mu}c^{e})\bar\omega^{bc}_{\nu}\right]\nonumber\\
&&-N^{ac}_{\mu\nu}\,D^{ab}_{\mu}\bar\omega^{bc}_{\nu}
-\left(\bar{M}^{ab}_{\mu\nu}M^{ab}_{\mu\nu}-\bar{N}^{ab}_{\mu\nu}N^{ab}_{\mu\nu}\right)\nonumber\\
&&+\bar{\lambda}^{ai}_{\alpha}(-\partial_{\mu}D^{ab}_{\mu})\lambda^{bi\alpha}+\bar\eta^{ai}_{\alpha}(-\partial_{\mu}D^{ab}_{\mu})\eta^{bi\alpha}
-gf^{abc}(\partial_{\mu}\bar\eta^{ai}_{\alpha})(D^{bd}_{\mu}c^{d})\lambda^{ci\alpha}\nonumber\\
&&+\bar{V}^{jk}_{\alpha\beta}\,\bar\psi^{i\alpha}g(T^{a})^{ij}\lambda^{ak\beta}
+\bar{U}^{jk}_{\alpha\beta}\left[ig^{2}(T^{a})^{ij}(T^{b})^{li}\bar\psi^{l\alpha}c^{b}\lambda^{ak\beta}
+\bar\psi^{i\alpha}g(T^{a})^{ij}\eta^{ak\beta}\right]\nonumber\\
&&+V^{ik}_{\alpha\beta}\left[\bar\lambda^{ak\beta}g(T^{a})^{ij}\psi^{j\alpha}-ig^{2}\bar\eta^{ak\beta}(T^{a})^{ij}(T^{b})^{jl}c^{b}\psi^{l\alpha}\right]
+U^{ik}_{\alpha\beta}\bar\eta^{ak\beta}g(T^{a})^{ij}\psi^{j\alpha}\nonumber\\
&&+\zeta m_{\psi}\,\left(\bar{V}^{ij}_{\alpha\beta}V^{ij}_{\alpha\beta}
-\bar{U}^{ij}_{\alpha\beta}U^{ij}_{\alpha\beta}\right)\biggr\}\,, 
\label{Sinv}
\end{eqnarray}
with
\begin{equation}
s S_{\mathrm{inv}} = 0 \;. \label{ssinv}
\end{equation}
The last term in expression \eqref{Sinv}, $\zeta m_{\psi}\,\left(\bar{V}^{ij}_{\alpha\beta}V^{ij}_{\alpha\beta}
-\bar{U}^{ij}_{\alpha\beta}U^{ij}_{\alpha\beta}\right)$,  is a vacuum term allowed  by power-counting, while $\zeta$  is a dimensionless coefficient. The starting action $S$ is recovered from  the invariant action  $S_{\mathrm{inv}}$ when the external sources attain a particular  value, usually called the physical value, {\it i.e.} 
\begin{equation}
S_{\mathrm{inv}} \Big|_{\rm phys \; value} = S 
\end{equation}  
where
\begin{equation}
M^{ab}_{\mu\nu}\Bigl|_{\mathrm{phys}}=\bar{M}^{ab}_{\mu\nu}\Bigl|_{\mathrm{phys}}=\gamma^{2}\delta^{ab}\delta_{\mu\nu}\,,\qquad
N^{ab}_{\mu\nu}\Bigl|_{\mathrm{phys}}=\bar{N}^{ab}_{\mu\nu}\Bigl|_{\mathrm{phys}}=0\,;
\end{equation}
\begin{equation}
V^{ij}_{\alpha\beta}\Bigl|_{\mathrm{phys}}=\bar{V}^{ij}_{\alpha\beta}\Bigl|_{\mathrm{phys}}=M^{3/2}\delta^{ij}\delta_{\alpha\beta}\,,\qquad
U^{ij}_{\alpha\beta}\Bigl|_{\mathrm{phys}}=\bar{U}^{ij}_{\alpha\beta}\Bigl|_{\mathrm{phys}}=0\,.
\end{equation}
As the action $S$ is obtained as a particular case of the extended action  $S_{\mathrm{inv}}$, renormalizability of $S_{\mathrm{inv}}$ will imply that  of $S$. We thus proceed by focussing on the action  $S_{\mathrm{inv}}$. \\\\In order to discuss the renormalizability of $S$ we notice  that the BRST transformations of the gauge, ghost and matter fields, eqs.\eqref{brstf},  are nonlinear. As such, we need to properly take into account the corresponding composite operators, a task which is achieved by introducing external invariant sources $(\Omega^{a}_{\mu}, L^{a},\bar{J}^{i\alpha},J^{i}_{\alpha})$ coupled to the nonlinear BRST transformations, {\it i.e.}
\begin{equation}
S_{\mathrm{ext}}=\int d^{4}x\,\left[\Omega^{a}_{\mu}\,sA^{a}_{\mu}+ L^{a}\,sc^{a}+\bar{J}^{i\alpha}\,s\psi^{i}_{\alpha}+(s\bar\psi^{i\alpha})J^{i}_{\alpha}\right]\,.
\end{equation}
Therefore, for  the complete starting classical action $\Sigma$ we obtain 
\begin{eqnarray}
\Sigma&=&S_{\mathrm{inv}}+S_{\mathrm{ext}}\nonumber\\
&=&\int d^{4}x\,\biggl\{\frac{1}{4}F^{a}_{\mu\nu}F^{a}_{\mu\nu}+
\bar{\psi}^{i\alpha}(\gamma_{\mu})_{\alpha\beta}D^{ij}_{\mu}\psi^{j\beta}
-m_{\psi}\bar{\psi}^{i}_{\alpha}\psi^{i\alpha}
+b^{a}\,\partial_{\mu}A^{a}_{\mu}
+\bar{c}^{a}\,\partial_{\mu}D^{ab}_{\mu}c^{b}
\nonumber\\
&&+\bar\varphi^{ac}_{\nu}\,\partial_{\mu}D^{ab}_{\mu}\varphi^{bc}_{\nu}
-\bar\omega^{ac}_{\nu}\,\partial_{\mu}D^{ab}_{\mu}\omega^{bc}_{\nu}
-gf^{abc}(\partial_{\mu}\bar\omega^{ae}_{\nu})(D^{bd}_{\mu}c^{d})\varphi^{ce}_{\nu}
-\bar{M}^{ac}_{\mu\nu}\,D^{ab}_{\mu}\varphi^{bc}_{\nu}\nonumber\\
&&
+\bar{N}^{ac}_{\mu\nu}\left[D^{ab}_{\mu}\omega^{bc}_{\nu}+gf^{abd}(D^{de}_{\mu}c^{e})\varphi^{bc}_{\nu}\right]
-M^{ac}_{\mu\nu}\left[D^{ab}_{\mu}\bar\varphi^{bc}_{\nu}+gf^{abd}(D^{de}_{\mu}c^{e})\bar\omega^{bc}_{\nu}\right]\nonumber\\
&&-N^{ac}_{\mu\nu}\,D^{ab}_{\mu}\bar\omega^{bc}_{\nu}
-\left(\bar{M}^{ab}_{\mu\nu}M^{ab}_{\mu\nu}-\bar{N}^{ab}_{\mu\nu}N^{ab}_{\mu\nu}\right)
-\bar{\lambda}^{ai}_{\alpha}\,\partial_{\mu}D^{ab}_{\mu}\lambda^{bi\alpha}-\bar\eta^{ai}_{\alpha}\,\partial_{\mu}D^{ab}_{\mu}\eta^{bi\alpha}\nonumber\\
&&-gf^{abc}(\partial_{\mu}\bar\eta^{ai}_{\alpha})(D^{bd}_{\mu}c^{d})\lambda^{ci\alpha}
+\bar{V}^{jk}_{\alpha\beta}\,\bar\psi^{i\alpha}g(T^{a})^{ij}\lambda^{ak\beta}
+\bar{U}^{jk}_{\alpha\beta}\left[\bar\psi^{i\alpha}g(T^{a})^{ij}\eta^{ak\beta}\right.\nonumber\\
&&\left.+ig^{2}(T^{a})^{ij}(T^{b})^{li}\bar\psi^{l\alpha}c^{b}\lambda^{ak\beta}\right]
+V^{ik}_{\alpha\beta}\left[\bar\lambda^{ak\beta}g(T^{a})^{ij}\psi^{j\alpha}-ig^{2}\bar\eta^{ak\beta}(T^{a})^{ij}(T^{b})^{jl}c^{b}\psi^{l\alpha}\right]\nonumber\\
&&+U^{ik}_{\alpha\beta}\bar\eta^{ak\beta}g(T^{a})^{ij}\psi^{j\alpha}
+\zeta m_{\psi}\,\left(\bar{V}^{ij}_{\alpha\beta}V^{ij}_{\alpha\beta}
-\bar{U}^{ij}_{\alpha\beta}U^{ij}_{\alpha\beta}\right)
-\Omega^{a}_{\mu}\,D^{ab}_{\mu}c^{b}+\frac{g}{2}f^{abc}L^{a}c^{b}c^{c} \nonumber\\
&&-\bar{J}^{i\alpha}\,ig(T^{a})^{ij}c^{a}\psi^{j}_{\alpha}
-ig\bar\psi^{j}_{\alpha}c^{a}(T^{a})^{ji}J^{i\alpha}\biggr\}\,.
\end{eqnarray}
Before proceeding  with the analysis of the renormalizability, it turns out  to be useful to introduce a multi-index notation. Following \cite{Zwanziger:1988jt,Zwanziger:1989mf,Zwanziger:1992qr}, we first introduce the multi-index 
$I\equiv\{a,\mu\}$, $I=1... 4(N^2-1)$, {\it i.e.}
\begin{eqnarray}
\left(\varphi^{ab}_{\mu},\bar\varphi^{ab}_{\mu},\omega^{ab}_{\mu},\bar\omega^{ab}_{\mu}\right)&\equiv&
\left(\varphi^{aI},\bar\varphi^{aI},\omega^{aI},\bar\omega^{aI}\right)\,,\nonumber\\
\left(M^{ab}_{\mu\nu},\bar{M}^{ab}_{\mu\nu},N^{ab}_{\mu\nu},\bar{N}^{ab}_{\mu\nu}\right)&\equiv&
\left(M^{aI}_{\mu},\bar{M}^{aI}_{\mu},N^{aI}_{\mu},\bar{N}^{aI}_{\mu}\right)\,.
\end{eqnarray}
As pointed out in \cite{Zwanziger:1988jt,Zwanziger:1989mf,Zwanziger:1992qr}, the possibility of introducing the multi-index $I\equiv\{a,\mu\}$
relies on an exact $U(f)$ symmetry, $f=4(N^{2}-1)$, {\it i.e.}
\begin{equation}
\mathcal{Q}^{ab}_{\mu\nu}(\Sigma)=0\,,
\label{U(f)}
\end{equation}
where
\begin{eqnarray}
\mathcal{Q}^{ab}_{\mu\nu}&\equiv&\int d^{4}x\,\biggl(\varphi^{ca}_{\mu}\frac{\delta}{\delta\varphi^{cb}_{\nu}}
-\bar\varphi^{cb}_{\nu}\frac{\delta}{\delta\bar\varphi^{ca}_{\mu}}
+\omega^{ca}_{\mu}\frac{\delta}{\delta\omega^{cb}_{\nu}}
-\omega^{cb}_{\nu}\frac{\delta}{\delta\omega^{ca}_{\mu}}\nonumber\\
&&+M^{ca}_{\sigma\mu}\frac{\delta}{\delta M^{cb}_{\sigma\nu}}
-\bar{M}^{cb}_{\sigma\nu}\frac{\delta}{\delta \bar{M}^{ca}_{\sigma\mu}}
+N^{ca}_{\sigma\mu}\frac{\delta}{\delta N^{cb}_{\sigma\nu}}
-\bar{N}^{cb}_{\sigma\nu}\frac{\delta}{\delta \bar{N}^{ca}_{\sigma\mu}}\biggr)\,.
\end{eqnarray}
Also, the trace of the operator $\mathcal{Q}^{ab}_{\mu\nu}$ defines the $q_f$-charge, displayed in the tables below. \\\\
Similarly, a second composite index $\hat{I}\equiv\{i,\alpha\}$ can be introduced,
\begin{eqnarray}
\left(\lambda^{ai\alpha},\bar\lambda^{ai\alpha},\eta^{ai\alpha},\bar\eta^{ai\alpha}\right)&\equiv&
\left(\lambda^{a\hat{I}},\bar\lambda^{a\hat{I}},\eta^{a\hat{I}},\bar\eta^{a\hat{I}}\right)\,,\nonumber\\
\left(V^{ij\alpha\beta},\bar{V}^{ij\alpha\beta},U^{ij\alpha\beta},\bar{U}^{ij\alpha\beta}\right)&\equiv&
\left(V^{i\alpha\hat{I}},\bar{V}^{i\alpha\hat{I}},U^{i\alpha\hat{I}},\bar{U}^{i\alpha\hat{I}}\right)\,,
\end{eqnarray}
due to a second  exact $U(\hat{f})$ symmetry,  $\hat{f}=4N$, {\it i.e.}
\begin{equation}
\hat{\mathcal{Q}}^{ij}_{\alpha\beta}(\Sigma)=0\,,
\label{U(hatf)}
\end{equation}
with
\begin{eqnarray}
\hat{\mathcal{Q}}^{ij}_{\alpha\beta}&\equiv&\int d^{4}x\,\biggl(\lambda^{ai}_{\alpha}\frac{\delta}{\delta\lambda^{aj\beta}}
-\bar\lambda^{aj}_{\beta}\frac{\delta}{\delta\bar\lambda^{ai\alpha}}
+\eta^{ai}_{\alpha}\frac{\delta}{\delta\eta^{aj\beta}}
-\bar\eta^{aj}_{\beta}\frac{\delta}{\delta\bar\eta^{ai\alpha}}\nonumber\\
&&+V^{ki}_{\gamma\alpha}\frac{\delta}{\delta V^{kj\beta}_{\gamma}}
-\bar{V}^{kj}_{\gamma\beta}\frac{\delta}{\delta \bar{V}^{ki\alpha}_{\gamma}}
+U^{ki}_{\gamma\alpha}\frac{\delta}{\delta U^{kj\beta}_{\gamma}}
-\bar{U}^{kj}_{\gamma\beta}\frac{\delta}{\delta \bar{U}^{ki\alpha}_{\gamma}}\biggr)\,.
\end{eqnarray}
The trace of the operator $\hat{\mathcal{Q}}^{ij}_{\alpha\beta}$ defines the $q_{\hat{f}}$-charge, also displayed in the tables below.  Making use of the multi-indices $(I, \hat{I})$,  for the complete action $\Sigma$ we get
\begin{eqnarray}
\Sigma&=&S_{\mathrm{inv}}+S_{\mathrm{ext}}\nonumber\\
&=&\int d^{4}x\,\biggl\{
\bar{\psi}^{i\alpha}(\gamma_{\mu})_{\alpha\beta}D^{ij}_{\mu}\psi^{j\beta}
-m_{\psi}\bar{\psi}^{i}_{\alpha}\psi^{i\alpha}
+\frac{1}{4}F^{a}_{\mu\nu}F^{a}_{\mu\nu}+b^{a}\,\partial_{\mu}A^{a}_{\mu}
+\bar{c}^{a}\,\partial_{\mu}D^{ab}_{\mu}c^{b}
\nonumber\\
&&+\bar\varphi^{aI}\,\partial_{\mu}D^{ab}_{\mu}\varphi^{bI}
-\bar\omega^{aI}\,\partial_{\mu}D^{ab}_{\mu}\omega^{bI}
-gf^{abc}(\partial_{\mu}\bar\omega^{aI})(D^{bd}_{\mu}c^{d})\varphi^{cI}
-\bar{M}^{aI}_{\mu}\,D^{ab}_{\mu}\varphi^{bI}\nonumber\\
&&
+\bar{N}^{aI}_{\mu}\left[D^{ab}_{\mu}\omega^{bI}+gf^{abc}(D^{cd}_{\mu}c^{d})\varphi^{bI}\right]
-M^{aI}_{\mu}\left[D^{ab}_{\mu}\bar\varphi^{bI}+gf^{abc}(D^{cd}_{\mu}c^{d})\bar\omega^{bI}\right]\nonumber\\
&&-N^{aI}_{\mu}\,D^{ab}_{\mu}\bar\omega^{bI}
-\left(\bar{M}^{aI}_{\mu}M^{aI}_{\mu}-\bar{N}^{aI}_{\mu}N^{aI}_{\mu}\right)
-\bar{\lambda}^{a}_{\hat{I}}\,\partial_{\mu}D^{ab}_{\mu}\lambda^{b\hat{I}}-\bar\eta^{a}_{\hat{I}}\,\partial_{\mu}D^{ab}_{\mu}\eta^{b\hat{I}}\nonumber\\
&&-gf^{abc}(\partial_{\mu}\bar\eta^{a}_{\hat{I}})(D^{bd}_{\mu}c^{d})\lambda^{c\hat{I}}
+\bar{V}^{j\hat{I}}_{\alpha}\,\bar\psi^{i\alpha}g(T^{a})^{ij}\lambda^{a}_{\hat{I}}
+\bar{U}^{j\hat{I}}_{\alpha}\left[\bar\psi^{i\alpha}g(T^{a})^{ij}\eta^{a}_{\hat{I}}\right.\nonumber\\
&&\left.+ig^{2}(T^{a})^{ij}(T^{b})^{ki}\bar\psi^{k\alpha}c^{b}\lambda^{a}_{\hat{I}}\right]
+V^{i\hat{I}}_{\alpha}\left[\bar\lambda^{a}_{\hat{I}}g(T^{a})^{ij}\psi^{j\alpha}
-ig^{2}\bar\eta^{a}_{\hat{I}}(T^{a})^{ij}(T^{b})^{jk}c^{b}\psi^{k\alpha}\right]\nonumber\\
&&+U^{i\hat{I}}_{\alpha}\bar\eta^{a}_{\hat{I}}g(T^{a})^{ij}\psi^{j\alpha}
+\zeta m_{\psi}\,\left(\bar{V}^{i\alpha\hat{I}}V^{i}_{\alpha\hat{I}}
-\bar{U}^{i\alpha\hat{I}}U^{i}_{\alpha\hat{I}}\right)
-\Omega^{a}_{\mu}\,D^{ab}_{\mu}c^{b}+\frac{g}{2}f^{abc}L^{a}c^{b}c^{c}\nonumber\\
&&-\bar{J}^{i\alpha}\,ig(T^{a})^{ij}c^{a}\psi^{j}_{\alpha}
-ig\bar\psi^{j}_{\alpha}c^{a}(T^{a})^{ji}J^{i\alpha}\biggr\}\,.
\label{actionfull}
\end{eqnarray}
Before ending this section, let us display the quantum numbers of all fields and external sources.  The nature of the fields/sources is denoted by ``B"  for bosonic fields/sources and by ``F" for anti-commuting variables. Also, the $e$-charge is the charge associated with a global $U(1)$ gauge invariance.
\begin{center}

\begin{tabular}{|l|c|c|c|c|c|c|c|c|c|c|c|c|c|c|}
\hline
\textsc{Fields}$\phantom{\Bigl|}$\!&$A^{a}_{\mu}$&$\psi^{i}_{\alpha}$&$\bar\psi^{i}_{\alpha}$&$c^{a}$&$\bar{c}^{a}$&$b^{a}$
&$\varphi^{aI}$&$\bar\varphi^{aI}$&$\omega^{aI}$&$\bar\omega^{aI}$
&$\lambda^{a\hat{I}}$&$\bar\lambda^{a\hat{I}}$&$\eta^{a\hat{I}}$&$\bar\eta^{a\hat{I}}$\\
\hline
\textsc{Dimension}&1&3/2&3/2&1&1&2&1&1&2&0&1&1&2&0\\
\textsc{Ghost number}&0&0&0&1&$-1$&0&0&0&1&$-1$&0&0&1&$-1$\\
$q_f$\textsc{-Charge}&0&0&0&0&0&0&1&$-1$&1&$-1$&0&$0$&0&0\\
$q_{\hat{f}}$\textsc{-Charge}&0&0&0&0&0&0&0&0&0&0&1&$-1$&1&$-1$\\
$e$\textsc{-Charge}&0&$1$&$-1$&0&0&0&0&0&0&0&1&$-1$&1&$-1$\\
\textsc{Nature}&B&F&F&F&F&B&B&B&F&F&F&F&B&B\\    
\hline

\end{tabular}

\begin{tabular}{|l|c|c|c|c|c|c|c|c|c|c|c|c|}
\hline
\textsc{Sources}$\phantom{\Bigl|}$\!&$\Omega^{a}_{\mu}$&$L^{a}$&$J^{i\alpha}$&$\bar{J}^{i\alpha}$&$M^{aI}_{\mu}$&$\bar{M}^{aI}_{\mu}$
&$N^{aI}_{\mu}$&$\bar{N}^{aI}_{\mu}$&$V^{i\alpha\hat{I}}$&$\bar{V}^{i\alpha\hat{I}}$
&$U^{i\alpha\hat{I}}$&$\bar{U}^{i\alpha\hat{I}}$\\
\hline
\textsc{Dimension}&2&2&3/2&3/2&2&2&3&1&3/2&3/2&5/2&1/2\\
\textsc{Ghost number}&$-1$&$-2$&$-1$&$-1$&$0$&0&1&$-1$&0&$0$&1&$-1$\\
$q_f$\textsc{-Charge}&0&0&0&0&1&$-1$&1&$-1$&0&0&0&0\\
$q_{\hat{f}}$\textsc{-Charge}&0&0&0&0&0&0&0&0&1&$-1$&1&$-1$\\
$e$\textsc{-Charge}&0&0&1&$-1$&0&0&0&0&0&0&0&0\\
\textsc{Nature}&F&B&B&B&B&B&F&F&B&B&F&F\\
\hline
\end{tabular}
\end{center}

\section{Ward identities}

In this section we derive the large set of Ward identities fulfilled by the complete action $(\Sigma)$. These Ward identities will be the starting point for the analysis of the algebraic characterization of the most general invariant counterterm. It is easily checked that $\Sigma$ obeys the following identities: 

\begin{itemize}
\item{The Slavnov-Taylor identity:
\begin{equation}
\mathcal{S}(\Sigma)= 0 \;,  \label{stid}
\end{equation}
with
\begin{eqnarray}
\mathcal{S}(\Sigma)&\equiv&\int d^{4}x\,\left(\frac{\delta\Sigma}{\delta\Omega^{a}_{\mu}}\frac{\delta\Sigma}{\delta A^{a}_{\mu}}
+\frac{\delta\Sigma}{\delta L^{a}}\frac{\delta\Sigma}{\delta c^{a}}
+\frac{\delta\Sigma}{\delta \bar{J}^{i}_{\alpha}}\frac{\delta\Sigma}{\delta\psi^{i\alpha}}
+\frac{\delta\Sigma}{\delta\bar\psi^{i}_{\alpha}}\frac{\delta\Sigma}{\delta J^{i\alpha}}
+b^{a}\frac{\delta\Sigma}{\delta\bar{c}^{a}}   \right. \nonumber\\
&&+\bar\varphi^{aI}\frac{\delta\Sigma}{\delta\bar\omega^{aI}}
+\omega^{aI}\frac{\delta\Sigma}{\delta\varphi^{aI}}
+\bar{M}^{aI}_{\mu}\frac{\delta\Sigma}{\delta\bar{N}^{aI}_{\mu}}
+N^{aI}_{\mu}\frac{\delta\Sigma}{\delta M^{aI}_{\mu}}\nonumber\\
&&  \left. +\bar\lambda^{i\hat{I}}\frac{\delta\Sigma}{\delta\bar\eta^{i\hat{I}}}
+\eta^{i\hat{I}}\frac{\delta\Sigma}{\delta\lambda^{i\hat{I}}}
+\bar{V}^{i\alpha\hat{I}}\frac{\delta\Sigma}{\delta\bar{U}^{i\alpha\hat{I}}}
+{U}^{i\alpha\hat{I}}\frac{\delta\Sigma}{\delta{V}^{i\alpha\hat{I}}}   \right)
\end{eqnarray}}

Let us also introduce, for further use, the so called  linearized Slavnov-Taylor operator \cite{Piguet:1995er} $\mathcal{B}_{\Sigma}$, defined as

\begin{eqnarray}
\mathcal{B}_{\Sigma}&=&\int d^{4}x\,\left( \frac{\delta\Sigma}{\delta\Omega^{a}_{\mu}}\frac{\delta}
{\delta A^{a}_{\mu}}+\frac{\delta\Sigma}{\delta A^{a}_{\mu}}\frac{\delta}{\delta\Omega^{a}_{\mu}}
+\frac{\delta\Sigma}{\delta L^{a}}\frac{\delta}{\delta c^{a}}+\frac{\delta\Sigma}{\delta
c^{a}}\frac{\delta}{\delta L^{a}}
+\frac{\delta\Sigma}{\delta \bar{J}^{i}_{\alpha}}\frac{\delta}{\delta\psi^{i\alpha}}
+\frac{\delta\Sigma}{\delta\psi^{i\alpha}}\frac{\delta}{\delta \bar{J}^{i}_{\alpha}}   \right. \nonumber\\
&&+\frac{\delta\Sigma}{\delta\bar\psi^{i}_{\alpha}}\frac{\delta}{\delta J^{i\alpha}}
+\frac{\delta\Sigma}{\delta J^{i\alpha}}\frac{\delta}{\delta\bar\psi^{i}_{\alpha}}
+b^{a}\frac{\delta}{\delta\bar{c}^{a}}
+\bar\varphi^{aI}\frac{\delta}{\delta\bar\omega^{aI}}
+\omega^{aI}\frac{\delta}{\delta\varphi^{aI}}
+\bar{M}^{aI}_{\mu}\frac{\delta}{\delta\bar{N}^{aI}_{\mu}}\nonumber\\
&& \left. +N^{aI}_{\mu}\frac{\delta}{\delta M^{aI}_{\mu}}
+\bar\lambda^{i\hat{I}}\frac{\delta}{\delta\bar\eta^{i\hat{I}}}
+\eta^{i\hat{I}}\frac{\delta}{\delta\lambda^{i\hat{I}}}
+\bar{V}^{i\alpha\hat{I}}\frac{\delta}{\delta\bar{U}^{i\alpha\hat{I}}}
+{U}^{i\alpha\hat{I}}\frac{\delta}{\delta{V}^{i\alpha\hat{I}}}\right) \,.
\end{eqnarray}

The operator $\mathcal{B}_{\Sigma}$ has the important property of being  nilpotent 
\begin{equation}
\mathcal{B}_{\Sigma} \mathcal{B}_{\Sigma} = 0 \;.
\end{equation}

\item{The Landau gauge-fixing condition \cite{Piguet:1995er}:
\begin{equation}
\frac{\delta\Sigma}{\delta b^{a}}=\partial_{\mu}A^{a}_{\mu}
\,.
\end{equation}}
\item{The anti-ghost equation \cite{Piguet:1995er}:
\begin{equation}
\frac{\delta\Sigma}{\delta\bar{c}^{a}}+\partial_{\mu}\frac{\delta\Sigma}{\delta\Omega^{a}_{\mu}}
=0  \;.
\end{equation}}
\item{The integrated ghost equation  \cite{Piguet:1995er}:
\begin{equation} 
G^{a}(\Sigma) = \Delta^a  \;,   \label{gheq}
\end{equation}
where 
\begin{eqnarray}
G^{a}(\Sigma)& \equiv & \int d^{4}x \left( \frac{\delta\Sigma}{\delta c^{a}}+gf^{abc}\left( 
\bar{c}^{b}\frac{\delta\Sigma}{\delta{b}^{c}}
+\bar{\omega}^{bI}\frac{\delta\Sigma}{\delta\bar{\varphi}^{cI}}
+{\varphi}^{bI}\frac{\delta\Sigma}{\delta{\omega}^{cI}}
+\bar{N}^{bI}_{\mu}\frac{\delta\Sigma}{\delta\bar{M}^{cI}_{\mu}}
+{M}^{bI}_{\mu}\frac{\delta\Sigma}{\delta{N}^{cI}_{\mu}}\right.  \right.\nonumber\\
&&
\left. \left. +\bar{\eta}^{b\hat{I}}\frac{\delta\Sigma}{\delta\bar{\lambda}^{c\hat{I}}}
+{\lambda}^{b\hat{I}}\frac{\delta\Sigma}{\delta{\eta}^{c\hat{I}}}\right)
+ig(T^{a})^{ij}{V}^{i\hat{I}}_{\alpha}\frac{\delta\Sigma}{\delta{U}^{j\hat{I}}_{\alpha}}
-ig(T^{a})^{ij}\bar{U}^{j\hat{I}}_{\alpha}\frac{\delta\Sigma}{\delta\bar{V}^{i\hat{I}}_{\alpha}}\right)\nonumber \;, 
\end{eqnarray}
and 
\begin{equation} 
\Delta^a = \int d^{4}x \left(gf^{abc}A_{\mu}^{c}\Omega^{b}_{\mu}
-gf^{abc}L^{b}c^{c}-\bar{J}^{i\alpha}ig(T^{a})^{ij}\psi^{j}_{\alpha}
+ig\bar{\psi}^{j}_{\alpha}(T^{a})^{ji}J^{i\alpha}\right)\,.
\end{equation}
Notice that the breaking term $\Delta^a$ is linear in the quantum fields. As such, it is a classical breaking, not affected by the quantum corrections  \cite{Piguet:1995er}. 
}
\item{The exact $U(f)$ symmetry \eqref{U(f)}, $f=4(N^2-1)$, here written in the multi-index notation 
\begin{equation} 
\mathcal{Q}^{IJ}(\Sigma) = 0  \;, \label{uf1} 
\end{equation} 
with 
\begin{eqnarray}
\mathcal{Q}^{IJ}(\Sigma)&\equiv&\int d^{4}x\,\Biggl(\varphi^{aI}\frac{\delta\Sigma}{\delta\varphi^{aJ}}
-\bar\varphi^{aJ}\frac{\delta\Sigma}{\delta\bar\varphi^{aI}}
+\omega^{aI}\frac{\delta\Sigma}{\delta\omega^{aJ}}
-\omega^{aJ}\frac{\delta\Sigma}{\delta\omega^{aI}}\nonumber\\
&&+M^{aI}_{\mu}\frac{\delta\Sigma}{\delta M^{aJ}_{\mu}}
-\bar{M}^{aJ}_{\mu}\frac{\delta\Sigma}{\delta \bar{M}^{aI}_{\mu}}
+N^{aI}_{\mu}\frac{\delta\Sigma}{\delta N^{aJ}_{\mu}}
-\bar{N}^{aJ}_{\mu}\frac{\delta\Sigma}{\delta \bar{N}^{aI}_{\mu}}\Biggr)\nonumber \;.
\end{eqnarray}

}
\item{The $U(\hat{f})$ symmetry, $\hat{f}=4N$:
\begin{equation}
{\hat{\mathcal{Q}}}^{\phantom{I}\hat{J}}_{\hat{I}}(\Sigma) = 0 \;, \label{hatf}
\end{equation}
where 
\begin{eqnarray}
{\hat{\mathcal{Q}}}^{\phantom{I}\hat{J}}_{\hat{I}}(\Sigma)&\equiv&\int d^{4}x\,\Biggl(
\lambda^{a}_{\hat{I}}\frac{\delta\Sigma}{\delta\lambda^{a}_{\hat{J}}}
-\bar\lambda^{a\hat{J}}\frac{\delta\Sigma}{\delta\bar\lambda^{a\hat{I}}}
+\eta^{a}_{\hat{I}}\frac{\delta\Sigma}{\delta\eta^{a}_{\hat{J}}}
-\bar\eta^{a\hat{J}}\frac{\delta\Sigma}{\delta\bar\eta^{a\hat{I}}}
\nonumber\\
&&
+V^{i\alpha}_{\,\,\,\,\,\hat{I}}\frac{\delta\Sigma}{\delta{V}^{i\alpha}_{\,\,\,\,\,\hat{J}}}
-\bar{V}^{i\alpha\hat{J}}\frac{\delta\Sigma}{\delta\bar{V}^{i\alpha\hat{I}}}
+U^{i\alpha}_{\,\,\,\,\,\hat{I}}\frac{\delta\Sigma}{\delta{U}^{i\alpha}_{\,\,\,\,\,\hat{J}}}
-\bar{U}^{i\alpha\hat{J}}\frac{\delta\Sigma}{\delta\bar{U}^{i\alpha\hat{I}}}
\Biggr)
\nonumber
\end{eqnarray}
}
\item{The $U(1)$ invariance:
\begin{equation}
\mathcal{N}_{e}(\Sigma) = 0 \;, \label{ne}
\end{equation}
with
\begin{eqnarray}
\mathcal{N}_{e}(\Sigma)&\equiv&\int d^{4}x\,\Biggl(
\psi^{i\alpha}\frac{\delta\Sigma}{\delta\psi^{i\alpha}}
-\bar\psi^{i\alpha}\frac{\delta\Sigma}{\delta\bar\psi^{i\alpha}}
+J^{i\alpha}\frac{\delta\Sigma}{\delta{J}^{i\alpha}}
-\bar{J}^{i\alpha}\frac{\delta\Sigma}{\delta\bar{J}^{i\alpha}}
+\lambda^{a\hat{I}}\frac{\delta\Sigma}{\delta\lambda^{a\hat{I}}}
-\bar\lambda^{a\hat{I}}\frac{\delta\Sigma}{\delta\bar\lambda^{a\hat{I}}}
\nonumber\\
&&
+\eta^{a\hat{I}}\frac{\delta\Sigma}{\delta\eta^{a\hat{I}}}
-\bar\eta^{a\hat{I}}\frac{\delta\Sigma}{\delta\bar\eta^{a\hat{I}}}
\Biggr)
\nonumber  \;.
\end{eqnarray}
This symmetry gives rise to a conserved charge,  called $e$ in the previous Tables. 
}.
\item{The ghost number Ward identity:
\begin{equation}
\mathcal{N_{\mathrm{ghost}}}(\Sigma)= 0 \;, \label{ghwid}
\end{equation}

\begin{eqnarray}
\mathcal{N_{\mathrm{ghost}}}(\Sigma)&\equiv&\int d^{4}x\,\Biggl(
c^{a}\frac{\delta\Sigma}{\delta{c}^{a}}
-\bar{c}^{a}\frac{\delta\Sigma}{\delta\bar{c}^{a}}
+\omega^{aI}\frac{\delta\Sigma}{\delta{\omega}^{aI}}
-\bar\omega^{aI}\frac{\delta\Sigma}{\delta\bar{\omega}^{aI}}
+\eta^{a\hat{I}}\frac{\delta\Sigma}{\delta{\eta}^{a\hat{I}}}
-\bar\eta^{a\hat{I}}\frac{\delta\Sigma}{\delta\bar{\eta}^{a\hat{I}}}
\nonumber\\
&&
+N^{aI}_{\mu}\frac{\delta\Sigma}{\delta{N}^{aI}_{\mu}}
-\bar{N}^{aI}_{\mu}\frac{\delta\Sigma}{\delta\bar{N}^{aI}_{\mu}}
+U^{i\hat{I}}_{\alpha}\frac{\delta\Sigma}{\delta{U}^{i\hat{I}}_{\alpha}}
-\bar{U}^{i\hat{I}}_{\alpha}\frac{\delta\Sigma}{\delta\bar{U}^{i\hat{I}}_{\alpha}}
-\Omega^{a}_{\mu}\frac{\delta\Sigma}{\delta\Omega^{a}_{\mu}}
-2L^{a}\frac{\delta\Sigma}{\delta L^{a}}
\nonumber\\
&&
-J^{i\alpha}\frac{\delta\Sigma}{\delta J^{i\alpha}}
-\bar{J}^{i\alpha}\frac{\delta\Sigma}{\delta\bar{J}^{i\alpha}}
\Biggr)
\nonumber \;.
\end{eqnarray}
}

\item{The linearly broken Ward identities:
\begin{equation}
\frac{\delta\Sigma}{\delta\bar{\varphi}^{aI}}+\partial_{\mu}\frac{\delta\Sigma}{\delta\bar{M}_{\mu}^{aI}}
=-gf^{abc}\bar{M}_{\mu}^{bI}A_{\mu}^{c}\,,
\end{equation}

\begin{equation}
\frac{\delta\Sigma}{\delta\omega^{aI}}+\partial_{\mu}\frac{\delta\Sigma}{\delta N_{\mu}^{aI}}-gf^{abc}\frac{\delta\Sigma}{\delta b^{a}}\bar{\omega}^{aI}
=-gf^{abc}A_{\mu}^{c}\bar{N}_{\mu}^{bI}\,,
\end{equation}

\begin{equation}
\frac{\delta\Sigma}{\delta\bar{\omega}^{aI}}+\partial_{\mu}\frac{\delta\Sigma}{\delta\bar{N}_{\mu}^{aI}}-gf^{abc}M_{\mu}^{bI}\frac{\delta\Sigma}{\delta\Omega_{\mu}^{a}}
 = gf^{abc}A_{\mu}^{c}\bar{N}_{\mu}^{bI}\,,
\end{equation}

\begin{equation}
\frac{\delta\Sigma}{\delta\varphi^{aI}}+\partial_{\mu}\frac{\delta\Sigma}{\delta M_{\mu}^{aI}}-gf^{abc}\left(\frac{\delta\Sigma}{\delta b^{c}}\bar{\varphi}^{bI}-\frac{\delta\Sigma}{\delta c^{c}}\bar{\omega}^{bI}+\bar{N}_{\mu}^{bI}\frac{\delta\Sigma}{\delta\Omega_{\mu}^{c}}\right)\nonumber \\
=-gf^{abc}\bar{M}_{\mu}^{bI}A_{\mu}^{c}\,,
\end{equation}

\begin{equation}
\int d^{4}x\left( \frac{\delta\Sigma}{\delta\eta^{a\hat{I}}}+gf^{abc}\bar{\eta}^{a}_{\hat{I}}\frac{\delta\Sigma}{\delta b^{a}}\right)
=\int d^{4}x\,g\left(T^{a}\right)^{ij}\bar{U}_{\alpha\hat{I}}^{j}\,\psi^{i\alpha}\,,
\end{equation}

\begin{equation}
\int d^{4}x\left( \frac{\delta\Sigma}{\delta\lambda^{a\hat{I}}}-g\left(T{}^{a}\right)^{ij}\bar{U}^{j\hat{I}}\frac{\delta\Sigma}{\delta J^{i\alpha}}+gf^{abc}\left( \frac{\delta\Sigma}{\delta b^{c}}\bar{\lambda}^{b\hat{I}}-\frac{\delta\Sigma}{\delta\bar{c}^{b}}\bar{\eta}^{c\hat{I}}\right) \right)
=-\int d^{4}x\,g\left(T^{a}\right)^{ij}\bar{V}_{\alpha\hat{I}}^{j}\,\bar{\psi}^{i\alpha}\,.
\end{equation}

\begin{equation}
\int d^{4}x\left( \frac{\delta\Sigma}{\delta\bar{\eta}_{\hat{I}}^{a}}-gV_{\alpha}^{i\hat{I}}\left(T^{a}\right)^{ij}\frac{\delta\Sigma}{\delta\bar{J}^{j\alpha}}\right) =\int d^{4}xgU_{\alpha}^{i\hat{I}}\left(T^{a}\right)^{ij}\psi^{j\alpha}
\end{equation}

\begin{equation}
\int d^{4}x\frac{\delta\Sigma}{\delta\bar{\lambda}_{\hat{I}}^{a}}=\int d^{4}xgV_{\alpha}^{i\hat{I}}\left(T^{a}\right)^{ij}\psi^{j\alpha}
\end{equation}

}
\item{the exact integrated Ward identities:
\begin{equation}
\int d^{4}x\left( 
c^{a}\frac{\delta\Sigma}{\delta\omega^{aI}}
+{\bar\omega}^{aI}\frac{\delta\Sigma}{\delta\bar{c}^{a}}
+\bar{N}^{aI}\frac{\delta\Sigma}{\delta\Omega^{a}_{\mu}}\right) =0\,,
\end{equation}

\begin{equation}
\int d^{4}x\left( 
c^{a}\frac{\delta\Sigma}{\delta\varphi^{aI}}
-{\bar\varphi}^{aI}\frac{\delta\Sigma}{\delta\bar{c}^{a}}
-\bar{M}^{aI}\frac{\delta\Sigma}{\delta\Omega^{a}_{\mu}}
+\frac{\delta\Sigma}{\delta\omega^{aI}}\frac{\delta\Sigma}{\delta{L}^{a}}\right) =0\,,
\end{equation}

\begin{equation}
\int d^{4}x\left( c^{a}\frac{\delta\Sigma}{\delta\eta^{a\hat{I}}}+\bar{\eta}^{a\hat{I}}\frac{\delta\Sigma}{\delta\bar{c}^{a}}-\bar{U}_{\mu}^{i\hat{I}}\frac{\delta\Sigma}{\delta J^{i\alpha}}\right) =0\,,
\end{equation}

\begin{equation}
\int d^{4}x\left(
c^{a}\frac{\delta\Sigma}{\delta\lambda^{a\hat{I}}}
-\bar\lambda^{a}_{\hat{I}}\frac{\delta\Sigma}{\delta\bar{c}^{a}}
-\frac{\delta\Sigma}{\delta\eta^{a\hat{I}}}\frac{\delta\Sigma}{\delta L^{a}}
-\bar{V}^{i}_{\alpha\hat{I}}\frac{\delta\Sigma}{\delta J^{i}_{\alpha}}
\right) =0\,.
\end{equation}

}
\end{itemize}

\section{Algebraic characterization of the most general invariant counterterm and renormalizability}

In order to determine the most general invariant counterterm which can be freely added to each order, we follow the algebraic renormalization framework  \cite{Piguet:1995er} and perturb  the complete action $\Sigma$ by adding an integrated local polynomial in the fields and sources with dimension four and vanishing ghost number, $\Sigma_{ct}$, and we require that the perturbed action, $(\Sigma + \varepsilon \Sigma_{ct})$, where $ \varepsilon $ is an infinitesimal expansion parameter, obeys the same Ward identities fulfilled by $\Sigma$ to the first order in the parameter $ \varepsilon$, obtaining the following constraints:  
\begin{equation}  
 \mathcal{B}_{\Sigma} \Sigma_{ct} = 0   \;, \label{bs} 
\end{equation}
\begin{equation} 
\frac{\delta\Sigma_{ct}}{\delta b^{a}}= 0 \;, \qquad     \frac{\delta\Sigma_{ct}}{\delta\bar{c}^{a}}+\partial_{\mu}\frac{\delta\Sigma_{ct}}{\delta\Omega^{a}_{\mu}} =0 \;, \label{bs2} 
\end{equation}
\begin{equation} 
G^{a}\Sigma_{ct} =0  \;, \label{bs3} 
\end{equation} 
\begin{equation} 
\mathcal{Q}^{IJ}\Sigma_{ct} =0 \;, \qquad  {\hat{\mathcal{Q}}}^{\phantom{I}\hat{J}}_{\hat{I}}\Sigma_{ct} = 0 \;, \label{bs4} 
\end{equation}
\begin{equation}
\mathcal{N}_{e}\Sigma_{ct} = 0 \;, \qquad  \mathcal{N_{\mathrm{ghost}}}\Sigma_{ct}= 0   \;, \label{bs5}
\end{equation}

\begin{equation}
\frac{\delta\Sigma_{ct} }{\delta\bar{\varphi}^{aI}}+\partial_{\mu}\frac{\delta\Sigma_{ct}}{\delta\bar{M}_{\mu}^{aI}}
=0 \;,  \qquad 
\frac{\delta\Sigma_{ct}}{\delta\omega^{aI}}+\partial_{\mu}\frac{\delta\Sigma_{ct}}{\delta N_{\mu}^{aI}}-gf^{abc}\frac{\delta\Sigma_{ct} }{\delta b^{a}}\bar{\omega}^{aI}
= 0 \;, \label{bs6} 
\end{equation}

\begin{equation}
\frac{\delta\Sigma_{ct}}{\delta\bar{\omega}^{aI}}+\partial_{\mu}\frac{\delta\Sigma_{ct} }{\delta\bar{N}_{\mu}^{aI}}-gf^{abc}M_{\mu}^{bI}\frac{\delta\Sigma_{ct} }{\delta\Omega_{\mu}^{a}}
 =0 \,, \qquad 
\int d^{4}x\left( \frac{\delta\Sigma_{ct} }{\delta\eta^{a\hat{I}}}+gf^{abc}\bar{\eta}^{a}_{\hat{I}}\frac{\delta\Sigma_{ct}}{\delta b^{a}}\right)
=0 \;, \label{bs7} 
\end{equation}

\begin{equation}
\frac{\delta\Sigma_{ct} }{\delta\varphi^{aI}}+\partial_{\mu}\frac{\delta\Sigma_{ct}}{\delta M_{\mu}^{aI}}-gf^{abc}\left(\frac{\delta\Sigma_{ct} }{\delta b^{c}}\bar{\varphi}^{bI}-\frac{\delta\Sigma_{ct}}{\delta c^{c}}\bar{\omega}^{bI}+\bar{N}_{\mu}^{bI}\frac{\delta\Sigma_{ct}}{\delta\Omega_{\mu}^{c}}\right)\nonumber \\
=0 \,,  \label{bs8} 
\end{equation}

\begin{equation}
\int d^{4}x\left( \frac{\delta\Sigma_{ct} }{\delta\lambda^{a\hat{I}}}-g\left(T{}^{a}\right)^{ij}\bar{U}^{j\hat{I}}\frac{\delta\Sigma_{ct}}{\delta J^{i\alpha}}+gf^{abc}\left(\frac{\delta\Sigma_{ct}}{\delta b^{c}}\bar{\lambda}^{b\hat{I}}-\frac{\delta\Sigma_{ct}}{\delta\bar{c}^{b}}\bar{\eta}^{c\hat{I}}\right)\right)
=0 \;, \label{bs9} 
\end{equation}

\begin{equation}
\int d^{4}x\left( \frac{\delta\Sigma_{ct} }{\delta\bar{\eta}_{\hat{I}}^{a}}-gV_{\alpha}^{i\hat{I}}\left(T^{a}\right)^{ij}\frac{\delta\Sigma_{ct}}{\delta\bar{J}^{j\alpha}}\right) =0 \;, \qquad 
\int d^{4}x\frac{\delta\Sigma_{ct} }{\delta\bar{\lambda}_{\hat{I}}^{a}}= 0\;, \label{bs10} 
\end{equation}

\begin{equation}
\int d^{4}x\left( 
c^{a}\frac{\delta\Sigma_{ct} }{\delta\omega^{aI}}
+{\bar\omega}^{aI}\frac{\delta\Sigma_{ct} }{\delta\bar{c}^{a}}
+\bar{N}^{aI}\frac{\delta\Sigma_{ct} }{\delta\Omega^{a}_{\mu}}\right) =0\,,   \label{bs11} 
\end{equation}

\begin{equation}
\int d^{4}x\left( 
c^{a}\frac{\delta\Sigma_{ct} }{\delta\varphi^{aI}}
-{\bar\varphi}^{aI}\frac{\delta\Sigma_{ct} }{\delta\bar{c}^{a}}
-\bar{M}^{aI}\frac{\delta\Sigma_{ct} }{\delta\Omega^{a}_{\mu}}
+\frac{\delta\Sigma }{\delta\omega^{aI}}\frac{\delta\Sigma_{ct} }{\delta{L}^{a}}
+\frac{\delta\Sigma }{\delta{L}^{a}}\frac{\delta\Sigma_{ct} }{\delta\omega^{aI}}\right) =0\,,   \label{bs12} 
\end{equation}

\begin{equation}
\int d^{4}x \left(  c^{a}\frac{\delta\Sigma_{ct} }{\delta\eta^{a\hat{I}}}+\bar{\eta}^{a\hat{I}}\frac{\delta\Sigma_{ct}}{\delta\bar{c}^{a}}-\bar{U}_{\mu}^{i\hat{I}}\frac{\delta\Sigma_{ct} }{\delta J^{i\alpha}}\right)=0\,,   \label{bs13} 
\end{equation}
and 
\begin{equation}
\int d^{4}x\left(
c^{a}\frac{\delta\Sigma_{ct}}{\delta\lambda^{a\hat{I}}}
-\bar\lambda^{a}_{\hat{I}}\frac{\delta\Sigma_{ct} }{\delta\bar{c}^{a}}
-\frac{\delta\Sigma}{\delta\eta^{a\hat{I}}}\frac{\delta\Sigma_{ct} }{\delta L^{a}}
-\frac{\delta\Sigma }{\delta L^{a}}\frac{\delta\Sigma_{ct}}{\delta\eta^{a\hat{I}}}
-\bar{V}^{i}_{\alpha\hat{I}}\frac{\delta\Sigma_{ct} }{\delta J^{i}_{\alpha}}
\right)=0\,.   \label{bs14} 
\end{equation}
The first condition, eq.\eqref{bs}, tells us that $\Sigma_{ct}$ belongs to the cohomolgy of the operator $\mathcal{B}_{\Sigma}$ in the space of the local integrated polynomials in the fields and external sources of dimension bounded by four. From the general results on the cohomolgy of Yang-Mills theories, see \cite{Piguet:1995er} and refs. therein, it follows that $\Sigma_{ct}$ can be parametrized as follows

\begin{equation}
\Sigma_{ct}=a_{0}S_{YM}+a_{1}m_{\psi}\int d^{4}x\; \bar{\psi}_{\alpha}^{i}\psi^{i\alpha}+\mathcal{B}_{\Sigma}\left(\Delta^{-1}\right)
\end{equation}
where $\Delta^{-1}$ is an integrated  polynomial of
dimension three and ghost number $-1$. The most general expression  for $\Delta^{-1}$ is given by 
\begin{eqnarray}
\Delta^{-1} & = & \int d^{4}x \biggl\{ a_{2}\Omega_{\mu}^{a}A_{\mu}^{a}+a_{3}L^{a}c^{a}+a_{4}\bar{J}^{i\alpha}\psi^{i\alpha}+a_{5}\bar{\psi}^{i\alpha}J^{i\alpha}
+a_{6}gf^{abc}\bar{N}_{\mu}^{aI}\varphi^{bI}A_{\mu}^{c}+a_{7}\bar{N}_{\mu}^{aI}\partial_{\mu}
\varphi^{aI}  \nonumber \\
&& +a_{8}gf^{abc}M_{\mu}^{aI}\bar{N}_{\mu}^{aI}
+a_{9}gf^{abc}M_{\mu}^{aI}A_{\mu}^{c}\bar{\omega}^{cI}
+a_{10}M_{\mu}^{aI}\partial_{\mu}\bar{\omega}^{bI}
+a_{11}A_{\mu}^{a}\partial_{\mu}\bar{c}^{a}\nonumber \\
&&
+a_{12}f^{abc}A_{\mu}^{a}\partial_{\mu}\varphi^{bI}\bar{\omega}^{cI}
+a_{13}gf^{abc}A_{\mu}^{a}\varphi^{bI}\partial_{\mu}\bar{\omega}^{cI}
+a_{14}gf^{abc}A_{\mu}^{a}\partial_{\mu}\lambda^{b\hat{I}}\bar{\eta}^{c\hat{I}}\nonumber \\
&&
+a_{15}gf^{abc}A_{\mu}^{a}\lambda^{b\hat{I}}\partial_{\mu}\bar{\eta}^{c\hat{I}}
+a_{16}\partial^{2}\varphi^{aI}\bar{\omega}^{aI}
+a_{17}\partial^{2}\lambda^{b\hat{I}}\bar{\eta}^{c\hat{I}}
+a_{18}gf^{abc}\bar{c}^{a}c^{b}\bar{c}^{c}\nonumber \\
&&
+a_{19}\bar{c}^{a}b^{a}
+a_{20}f^{abc}\bar{c}^{a}\varphi^{bI}\bar{\varphi}^{cI}
+a_{21}gf^{abc}\bar{c}^{a}\lambda^{b\hat{I}}\bar{\lambda}^{c\hat{I}}
+a_{22}gf^{abc}\bar{c}^{a}\omega^{bI}\bar{\omega}^{cI}\nonumber \\
&&
+a_{23}gf^{abc}\bar{c}^{a}\eta^{b\hat{I}}\bar{\eta}^{c\hat{I}}
+a_{24}gf^{abc}b^{a}\varphi^{bI}\bar{\omega}^{cI}
+a_{25}gf^{abc}b^{a}\lambda^{b\hat{I}}\bar{\eta}^{c\hat{I}}
+a_{26}\zeta m_{\psi}V^{i\alpha\hat{I}}\bar{U}^{i\alpha\hat{I}}\nonumber \\
&&
+a_{27}m_{\psi}^{2}\varphi^{aI}\bar{\omega}^{aI}
+a_{28}m_{\psi}^{2}\lambda^{a\hat{I}}\bar{\eta}^{a\hat{I}}
+a_{29}g(T^a)^{ij}\bar{\psi}^{i\alpha}\lambda^{a\hat{I}}\bar{U}^{j\alpha\hat{I}}
+a_{30}g(T^{a})^{ij} \psi^{j\alpha}\bar\eta^{a\hat{I}} V^{i\alpha\hat{I}}\nonumber \\
&& 
+a_{31} (\gamma_\mu)_{\alpha\beta} (T^a)^{ij} A_{\mu}^{a}V^{i\alpha\hat{I}}\bar{U}^{j\beta\hat{I}}
+a_{32} (\gamma_\mu)_{\alpha\beta} (\partial_{\mu}V^{i\alpha\hat{I}})\bar{U}^{i\beta\hat{I}}
+\mathbb{C}_{1}^{abcd}A_{\mu}^{a}A_{\mu}^{b}\varphi^{cI}\bar{\omega}^{dI}
+\mathbb{C}_{2}^{abcd}A_{\mu}^{a}A_{\mu}^{b}\lambda^{c\hat{I}}\bar{\eta}^{d\hat{I}}\nonumber \\
&&
+\mathbb{C}_{3}^{abcdeIJLM}c^{a}\varphi^{bI}\varphi^{cJ}\bar{\omega}^{dL}\bar{\omega}^{eM}
+\mathbb{C}_{4}^{abcde\hat{I}\hat{J}\hat{L}\hat{M}}c^{a}\lambda^{b\hat{I}}\lambda^{c\hat{J}}\bar{\eta}^{d\hat{L}}\bar{\eta}^{e\hat{M}}
+\mathbb{C}_{5}^{abcd}\varphi^{aI}\bar{\omega}^{bI}\lambda^{c\hat{I}}\bar{\lambda}^{d\hat{I}}\nonumber \\
&&
+\mathbb{C}_{6}^{abcd}\varphi^{aI}\bar{\varphi}^{bI}\lambda^{c\hat{I}}\bar{\eta}^{d\hat{I}}
+\mathbb{C}_{7}^{abcdIJLM}\varphi^{aI}\bar{\omega}^{bI}\omega^{cL}\bar{\omega}^{dM}
+\mathbb{C}_{8}^{abcdIJLM}\varphi^{aI}\bar{\varphi}^{bI}\varphi^{cL}\bar{\omega}^{dM}\nonumber \\
&& 
+\mathbb{C}_{9}^{abcd\hat{I}\hat{J}\hat{L}\hat{M}}\lambda^{a\hat{I}}\bar{\eta}^{b\hat{J}}\eta^{c\hat{L}}\bar{\eta}^{d\hat{M}}
+\mathbb{C}_{10}^{abcd\hat{I}\hat{J}\hat{L}\hat{M}}\lambda^{a\hat{I}}\bar{\lambda}^{b\hat{J}}\lambda^{c\hat{L}}\bar{\eta}^{d\hat{M}} \biggr\} \;.
\end{eqnarray}
After imposition of the remaining conditions, els.\eqref{bs2}-\eqref{bs14}, and after a rather long  algebra, it turns out that  the only non-vanishing 
coefficients  are\footnote{After application of the linearized BRST operator, the pure external source term $a_{32} (\gamma_\mu)_{\alpha\beta} (\partial_{\mu}V^{i\alpha\hat{I}})\bar{U}^{i\beta\hat{I}}$ gives rise to $a_{32} (\gamma_\mu)_{\alpha\beta} \left( (\partial_{\mu}U^{i\alpha\hat{I}})\bar{U}^{i\beta\hat{I}}+ (\partial_{\mu}V^{i\alpha\hat{I}})\bar{V}^{i\beta\hat{I}}\right)$, which identically vanishes when the sources attain their physical values. Therefore, from now on we shall set $a_{32}=0$.}
\begin{equation}
a_{2}=-a_{6}=a_{7}=a_{8}=a_{9}=a_{10}=a_{11}=a_{13}=a_{15}=-a_{16}=a_{17}\ne 0\,,
\end{equation}
\begin{equation}
a_{4}=a_{5}=a_{29}=a_{30}\ne 0\,,\quad\quad a_{26}\ne 0\,.
\end{equation}
Thus, for the most general invariant local counterterm one finds  
\begin{eqnarray}
\Sigma_{ct} & = & \int d^{4}x \biggl\{ a_{0}F_{\mu\nu}^{a}F_{\mu\nu}^{a}
+a_{1}m_{\psi}\bar{\psi}_{\alpha}^{i}\psi^{i\alpha}+a_{2}\left[\frac{\delta S_{YM}}{\delta A_{\mu}^{a}}A_{\mu}^{a}\right.
+\partial_{\mu}\bar{c}^{a}\partial_{\mu}c^{a}+\Omega_{\mu}^{a}\partial_{\mu}c^{a}\nonumber \\
 &  & +gf^{abc}\left(\partial_{\mu}c^{a}\bar{N}_{\mu}^{bI}\varphi^{cI}+\partial_{\mu}c^{a}M_{\mu}^{bI}\bar{\omega}^{cI}
-\partial_{\mu}c^{a}\varphi^{bI}\partial_{\mu}\bar{\omega}^{cI}
-\partial_{\mu}c^{a}\lambda^{b\hat{I}}\partial_{\mu}\bar{\eta}^{c\hat{I}}\right)\nonumber \\
 &  & +\bar{\omega}^{aI}\partial^{2}\omega^{aI}-\bar{\varphi}^{aI}\partial^{2}\varphi^{aI}
+\eta^{aI}\partial^{2}\bar{\eta}^{aI}-\lambda^{aI}\partial^{2}\bar{\lambda}^{aI}
+\bar{M}_{\mu}^{aI}\partial_{\mu}\varphi^{aI}+\omega^{aI}\partial_{\mu}\bar{N}_{\mu}^{aI}\nonumber \\
 &  & +N_{\mu}^{aI}\partial_{\mu}\bar{\omega}^{aI}-\bar{\varphi}^{aI}\partial_{\mu}M_{\mu}^{aI}
+\left.\bar{M}_{\mu}^{aI}M_{\mu}^{aI}-\bar{N}_{\mu}^{aI}N_{\mu}^{aI}\right]
+a_{26}\zeta m_{\psi}\left(\bar{V}^{i\alpha\hat{I}}V^{i\alpha\hat{I}}
-\bar{U}^{i\alpha\hat{I}}U^{i\alpha\hat{I}}\right)\nonumber \\
 &  & +a_{4}\left[-2\bar{\psi}^{i\alpha}\left(\gamma_{\mu}\right)_{\alpha\beta}D_{\mu}^{ij}\psi^{j\beta}
+2m_{\psi}\bar{\psi}_{\alpha}^{i}\psi^{i\alpha}\right]     \biggr\}
\end{eqnarray}
To complete the analysis of the algebraic renormalization of the model, we need to show that the counterterm $\Sigma_{ct}$ can be reabsorbed into the starting action $\Sigma$ through a redefinition of the fields $\{\phi \}$, $\phi=(A,c,{\bar c}, b, \varphi, {\bar \varphi}, \omega, {\bar \omega}, \psi, {\bar \psi}, \lambda, {\bar \lambda}, \eta, {\bar \eta})$,   sources $\{ S \}$, $S=(\Omega, L, J, {\bar J}, {\bar N}, {\bar M}, N, M, {\bar U}, {\bar V}, U, V)$, and parameters $\tau$, $\tau=(g, \zeta, m_{\psi})$, namely 
\begin{equation}
\label{ration}
\Sigma(\phi,S,\tau) +  \varepsilon \Sigma_{ct}(\phi,S,\tau)  = \Sigma(\phi_0,S_0,\tau_0) + O( \varepsilon^2) \;, 
\end{equation}
where $(\phi_0, S_0, \tau_0)$ stand for the so-called bare fields, sources and parameters.

\begin{equation}
\phi_{0}=Z^{1/2}_{\phi}\,\phi  \qquad\;,   \qquad
S_{0}=Z_{S}\,S\,,  \qquad \tau_0 = Z_{\tau} \tau   \;, \label{renormfs}
\end{equation}
By direct inspection, for the renormalization factors we find 
\begin{equation}
Z_{A}^{1/2}=1+\varepsilon\left(\frac{a_{0}}{2}+a_{2}\right) \;, \qquad 
Z_{g}=1-\frac{\varepsilon a_{0}}{2}  \;, 
\end{equation}
\begin{equation}
Z_{\bar{\psi}}^{1/2}=Z_{\psi}^{1/2}=1-\varepsilon a_{4} \;, \qquad 
Z_{m_{\psi}}^{1/2}=1-\varepsilon a_{1} \;, 
\end{equation}
\begin{equation}
Z_{b}^{1/2}=Z_{A}^{-1/2} \;, \qquad 
Z_{\bar{c}}^{1/2}=Z_{c}^{1/2}=Z_{A}^{-1/4}Z_{g}^{-1/2} \;, 
\end{equation}
\begin{equation}
Z_{\varphi}^{1/2}=Z_{\bar{\varphi}}^{1/2}=Z_{A}^{-1/4}Z_{g}^{-1/2}\;, \qquad 
Z_{\bar{\omega}}^{1/2}=Z_{g}^{-1} \;, 
\end{equation}
\begin{equation}
Z_{\omega}^{1/2}=Z_{A}^{-1/2}\;, \qquad 
Z_{\lambda}^{1/2}=Z_{\bar{\lambda}}^{1/2}=Z_{A}^{-1/4}Z_{g}^{-1/2} \;, 
\end{equation}
\begin{equation}
Z_{\bar{\eta}}^{1/2}=Z_{g}^{-1}\;, \qquad 
Z_{\eta}^{1/2}=Z_{A}^{-1/2} \;, 
\end{equation}
\begin{equation}
Z_{\bar{M}}=Z_{M}=Z_{A}^{-1/4}Z_{g}^{-1/2}\;, \qquad 
Z_{\bar{N}}=Z_{g}^{-1}  \;, 
\end{equation}
\begin{equation}
Z_{N}=Z_{A}^{-1/2}\;, \qquad 
Z_{\bar{V}}=Z_{V}=Z_{\psi}^{-1/2}Z_{g}^{-1/2}Z_{A}^{1/4} \;, 
\end{equation}
\begin{equation}
Z_{\bar{U}}=Z_{\psi}^{-1/2}Z_{g}^{-1}Z_{A}^{1/2}\;, \qquad 
Z_{U}=Z_{\psi}^{-1/2} \;, 
\end{equation}
\begin{equation}
Z_{\zeta}=\left(1+\varepsilon a_{26}\right)Z_{m_{\psi}}^{-1}Z_{\psi}Z_{g}Z_{A}^{-1/2} \;. 
\end{equation}
\begin{equation} 
Z_{\Omega} = Z_A^{-1/2} Z_c^{-1/2} Z_g^{-1}  \;, \qquad Z_L = Z_A^{1/2} \;, \qquad Z_J=Z_{\bar J} = Z_g^{-1} Z_c^{-1/2} Z_{\psi}^{-1/2} \;. 
\end{equation} 
This ends the analysis of the all orders  algebraic renormalization of the action $\Sigma$, and thus of the starting action $S$, eq.\eqref{starting}.

\section{Conclusion}
In this work we have pursued the investigation started in \cite{Capri:2014bsa}, where the coupling between the inverse of the Faddeev-Popov operator and quark matter fields has been introduced through the operators 
\begin{eqnarray}
\mathcal{R}_{\psi}^{ai}(x)&:=&g\int d^{4}z (\mathcal{M}^{-1})^{ab}(x,z)(T^{b})^{ij}\psi^{j}(z)\,,\nonumber\\
\bar{\mathcal{R}}_{\psi}^{ai}(x)&:=&g\int d^{4}z (\mathcal{M}^{-1})^{ab}(x,z)\bar{\psi}^{j}(z)(T^{b})^{ji}\,,
\label{rpsi}
\end{eqnarray}
giving rise to a non-trivial correlation function  \cite{Capri:2014bsa}
\begin{equation}
Q^{abij}_{\psi}(x-y)=\left\langle {\mathcal{R}}^{ai}_{\psi}(x)\bar{\mathcal{R}}^{bj}_{\psi}(y)\right\rangle\,.     \label{qpsi}
\end{equation}
This correlation function can be directly studied in lattice numerical simulations, as done recently in the case of gluons \cite{Cucchieri:2014via}. \\\\As shown in \cite{Capri:2014bsa}, the correlation function \eqref{qpsi}  can be obtained from a theory constructed by adding to the usual matter action a term similar to Zwanzige's horizon function \cite{Zwanziger:1988jt,Zwanziger:1989mf,Zwanziger:1992qr}, namely
\begin{equation}
S_{\mathrm{matter}}\to S_{\mathrm{matter}}+M^{3}H_{\mathrm{matter}}\,,
\label{matter_action_psi}
\end{equation}
where, in complete analogy with the horizon function for the gluon sector \cite{Zwanziger:1988jt,Zwanziger:1989mf,Zwanziger:1992qr},  $H_{\mathrm{matter}}$ is given by
\begin{equation}
H_{\mathrm{matter}}=-g^2\int d^4xd^4y\,\bar{\psi}^{\,i}(x)(T^a)^{ij}(\mathcal{M}^{-1})^{ab}(x,y)(T^b)^{jk}\psi^{\,k}(y)  \;.
\label{horizon_function_psi}
\end{equation}
with the mass parameter $M$ playing a role similar to that of the Gribov parameter $\gamma$. Remarkably, the introduction of such a non-local term gives rise to a quark propagator of the kind 
\begin{equation}
\left\langle{\psi}^{i}(p)\bar{\psi}^{j}(-p)\right\rangle=\frac{-ip_{\mu}\gamma_{\mu}+\mathcal{A}(p^{2})}{p^{2}+\mathcal{A}^2(p^{2})}\delta^{ij}\,,
\label{psi_propagator}
\end{equation}
with the quark mass function $\mathcal{A}(p^{2})$  given by
\begin{equation}
\mathcal{A}(p^{2})=m_{\psi}+g^{2}\left(\frac{N^{2}-1}{2N}\right)\,\frac{M^{3}}{p^{2}+\mu^{2}_{\psi}}\,,   \label{aq1}
\end{equation}
This propagator fits very well the available lattice numerical  data, see  \cite{Parappilly:2005ei} and the discussion  in \cite{Capri:2014bsa}. \\\\In the present paper we have been able to show that, despite its non-locality, expression \eqref{horizon_function_psi} can be cast in local form by means of a set of suitable localizing fields. Moreover, when cast in local form, the resulting action fulfils a large set of Ward identities which have enabled us to prove that the theory is multiplicative renormalizable to all orders. This is a non-trivial result which supports the idea that the Faddeev-Popov operator couples in a universal way to both gauge and quark fields, as expressed by equations \eqref{R} and \eqref{rpsi}. \\\\This coupling allows us to reproduce from a local and renormalizable action the behaviour of the gluon and quark propagators observed in numerical simulations, reinforcing the belief that the inverse of the Faddeev-Popov operator  plays an important role in the infrared dynamics of confining Yang-Mills theories.

\section*{Acknowledgments}
The Conselho Nacional de Desenvolvimento Cient\'{\i}fico e
Tecnol\'{o}gico (CNPq-Brazil), the Faperj, Funda{\c{c}}{\~{a}}o de
Amparo {\`{a}} Pesquisa do Estado do Rio de Janeiro,  the
Coordena{\c{c}}{\~{a}}o de Aperfei{\c{c}}oamento de Pessoal de
N{\'{\i}}vel Superior (CAPES)  are gratefully acknowledged.

\end{document}